\documentclass[final,english]{bullsrsl}[2022/06/15]

\usepackage{subfigure}
\usepackage[latin1]{inputenc}
\usepackage{adjustbox}
\usepackage[T1]{fontenc}
\usepackage{rotating}
\usepackage{natbib} 
\usepackage{graphicx}
\newcommand\LiU{Li\`{e}ge}
\newcommand\third{$3^{\rm rd}$}
\newcommand\arcdeg{\mbox{$^\circ$}}%
\newcommand\arcmin{\mbox{$^\prime$}}%
\newcommand\arcsec{\mbox{$^{\prime\prime}$}}%
\begin{document}
\title{Indo-Belgian co-operation in Astrophysics: From inception to future prospects}

\author{Ram}{Sagar}
\affiliation[]{ Indian Institute of Astrophysics, Bangalore 560034 and Aryabhatta Research Institute of Observational sciencES (ARIES), Manora Peak, Nainital, 263001, India }
\correspondance{ram\_sagar0@yahoo.co.in}
\date{15th May 2023}
\maketitle 
\begin{abstract}

In this manuscript, an overview of the accomplishments of the Indo-Belgian co-operation is presented in the current era of multi-wavelength global astronomy. About two decades ago, in the field of astronomy and astrophysics, academicians from India and Belgium embarked on formal interaction and collaboration. The Belgo-Indian Network for Astronomy \& astrophysics (BINA), initiated in 2014, has been very productive and its activities have set a landmark for Indo-Belgian co-operation. Under this program, three international workshops were conducted. Several exchange work visits were also made among the astronomers of the two countries. Since the necessary foundation work has already been done, continuation of the BINA activities in future is strongly recommended.

\end{abstract}


\section{Introduction}
\label{intro}
The Indo-Belgian Research and Technology co-operation germinated during 2004-2005 and became operative on November 3, 2006, when the Department of Science and Technology (DST; Govt. of India) and the Belgian Federal Science Policy Office (BELSPO; Govt. of Belgium) signed a Memorandum of Understanding (MoU) in New Delhi. This MoU covers a wide range of scientific fields including physics and astrophysics, one of the key areas for long-term co-operation. A joint committee (JC) consisting of officials and experts from both countries was constituted for periodic monitoring and smooth implementation of the collaborative projects. The first Indo-Belgium JC meeting was held in Brussels during June 23-28, 2007. The Indian Science delegation was led by Prof. T. Ramasamy, who was the secretary of DST at that time. To strengthen the Indo-Belgian collaborations in the area of astronomy and astrophysics, author in the role of director, ARIES visited a few Belgian astronomical research institutions from May 11 to 15, 2012. On May 14, 2012, he also participated in and delivered an invited talk during the FNRS Contact Group  Astronomie \& Astrophysique meeting held on astronomy day of the Royal Observatory of Belgium (ROB) at the planetarium in Brussels. During such meetings, an open discussion regarding the future of Belgian astronomy takes place. 

\begin{figure}[h!]
\centering
\includegraphics[width=14cm]{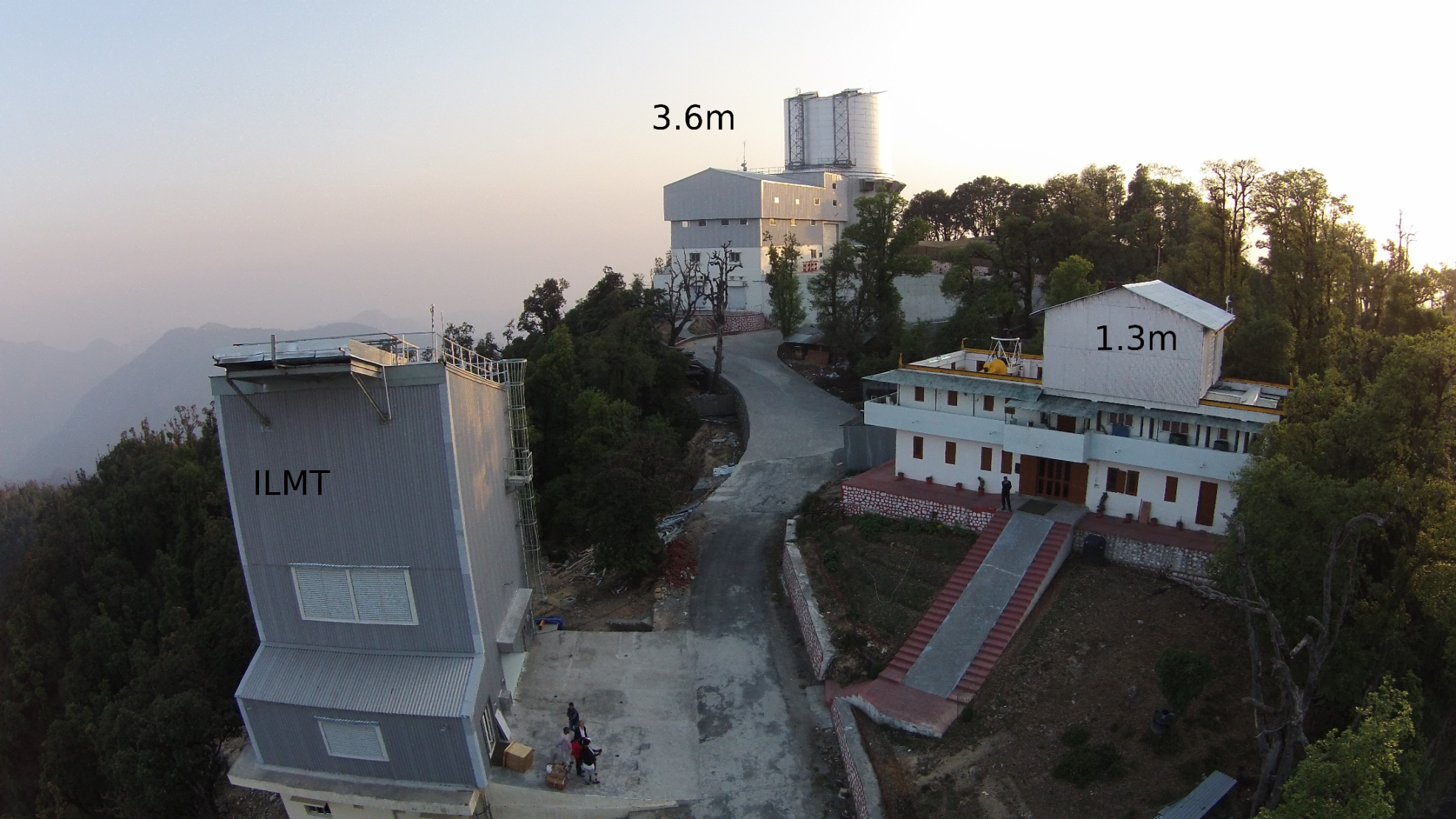}
\bigskip
\caption {An aerial view of the Devasthal observatory located in the central Himalayan region of Nainital, Uttarakhand (Longitude=$79\arcdeg 41\arcmin 04\arcsec$ E, Latitude=$29\arcdeg 21\arcmin 40\arcsec$ N and Altitude = $2424 \pm 4$ m). The housing of all the optical telescopes, marked with their sizes, are shown.} 
\label{fig:DevObs}
\end{figure}

  Three optical telescopes, namely the 1.3-m Devasthal Fast Optical Telescope (DFOT), 3.6-m Devasthal Optical Telescope (DOT) and 4-m International Liquid Mirror Telescope (ILMT), have been successfully installed at the Devasthal observatory. Its growth path has been chronicled recently by \citet{sagar2023}. The observatory is managed and operated by the Aryabhatta Research Institute of observational sciencES (ARIES) in Nainital (India), which is an autonomous research institution under DST \citep{2022InJHS...57..227S}. The Belgians are partners in both 3.6-m DOT and 4-m ILMT. An aerial view of the Devasthal observatory is shown in Fig.\,\ref{fig:DevObs}. The building with the sliding roof towards the bottom right is the 1.3-m DFOT \citep{2011CSci..101.1020S}, the rectangular building towards the bottom left is the 4-m ILMT and the topmost is the 3.6-m DOT dome and extension buildings.
  
In 2014, the Belgo-Indian Network for Astronomy and astrophysics (BINA) was created under the above-mentioned MoU with an aim to foster collaborations in space research between different institutions of both countries. The BINA links researchers of the 13 Indian and 6 Belgian institutions \citep{2019BSRSL..88...19J}. Indian institutions, led by ARIES, Nainital (PI : Dr. Santosh Joshi), are located in different parts of the country while both federal and regional organizations of  Belgium are  participating in the activities of BINA under the leadership of  the ROB, Brussels (PI : Dr. Peter De Cat). Thus, a good number of researchers from both countries are collaborating under BINA and organizing both joint workshops and mutual work visits. Such activities have strengthened research in the areas of the solar system, galactic and extra-galactic astronomy and have also contributed to the development of back-end instruments for Indo-Belgian telescopes. 
 
 The next section describes the participation of the Belgians in the development of the Devasthal observatory. The main accomplishments of the Indo-Belgian cooperation, including BINA, are summarised in Sec. \ref{accomp}. Finally, in the last section, we have listed the scientific potential of the BINA along with the possibilities of future collaborations. 
   
\section{Indo-Belgium partnership in both 4-m class optical telescopes }  
\label{partner}
\begin{figure}[h!]
\centering
\includegraphics[width=12cm]{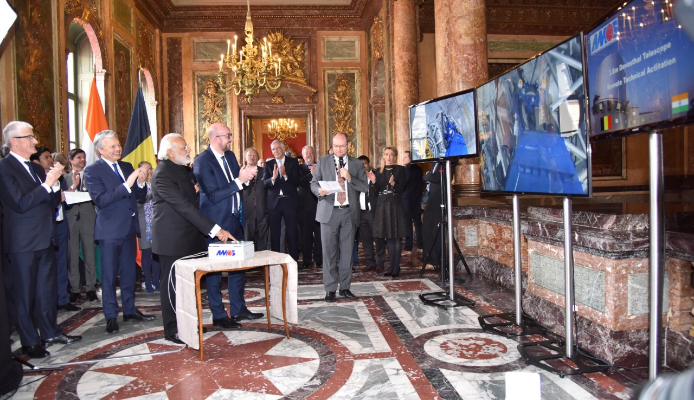}
\bigskip
\caption {A photograph of technical activation of the 3.6-m DOT by the premiers of India (Shri Narendra Modi) and Belgium (Mr. Charles Michel) on March 30, 2016.} 
\label{fig:acti_dot}
\end{figure}

Belgium is a partner in both 3.6-m DOT and 4-m ILMT, India's largest 4-m class optical observational facilities. On March 27, 2007, ARIES awarded the contract for the supply and installation of a 3.6-m aperture-size modern optical telescope at Devasthal to Advanced Mechanical and Optical Systems (AMOS), a globally well-known Belgian company for the delivery of precision astronomical instruments. In the year 2008, BELSPO contributed 2 million Euros in cash to this project. In return, Belgian astronomers are assured of 7\% observing time on this modern observing facility. This Indo-Belgian telescope was technically activated on March 30, 2016 (Fig.\,\ref{fig:acti_dot}), jointly by the premiers of India, Shri Narendra Modi and of Belgium, Mr. Charles Michel \citep{2017CSci..113..682O, 2018BSRSL..87...29K, 2019CSci..117..365S}. 

\begin{figure}[h!]
\centering
\includegraphics[width=12cm]{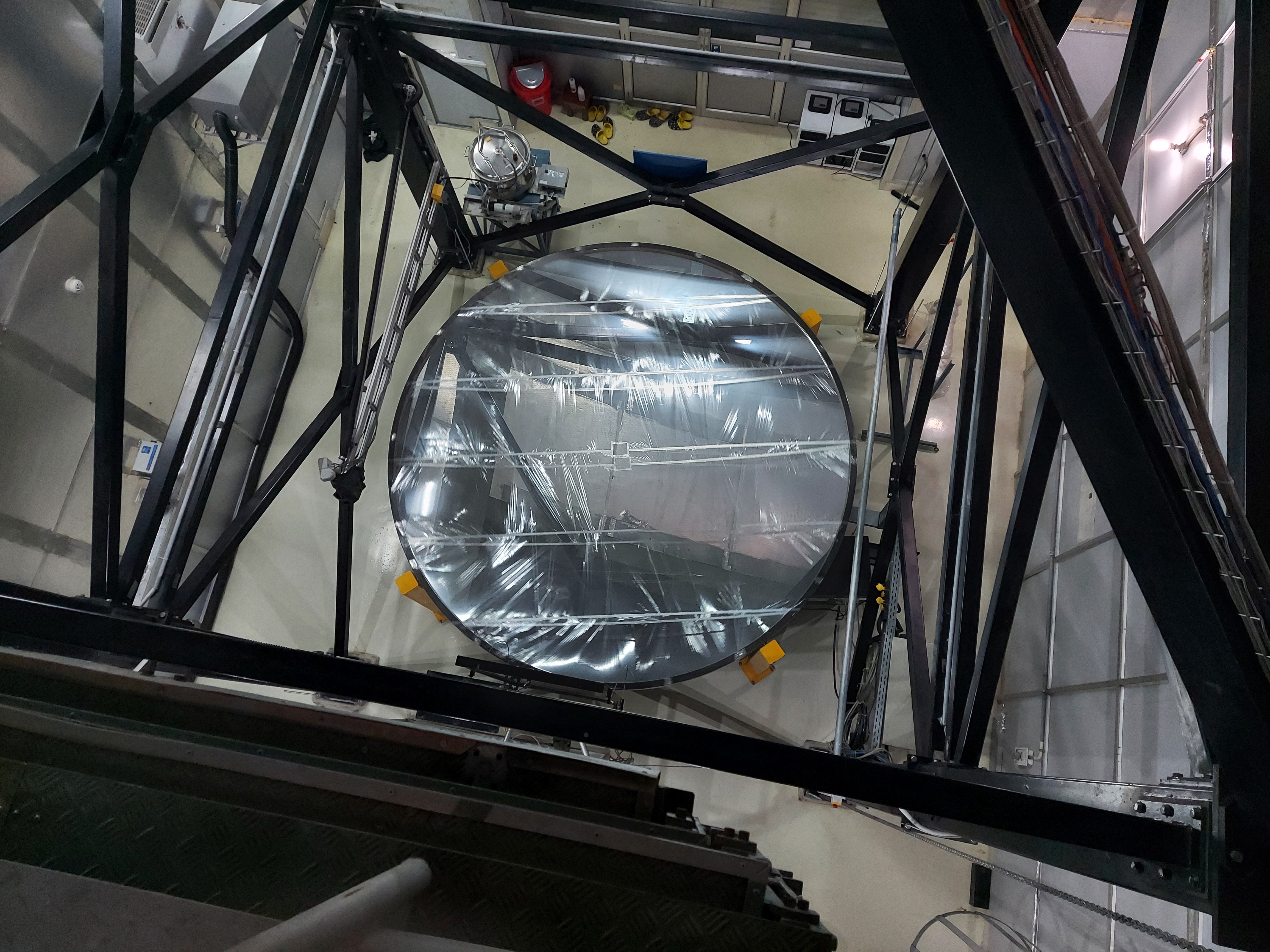}
\bigskip
\caption{ A top view of the ILMT showing the liquid mercury mirror covered by a mylar film.}
\label{fig:ILMT}
\end{figure}

\begin{figure}[h!]
\centering
\includegraphics[width=12cm]{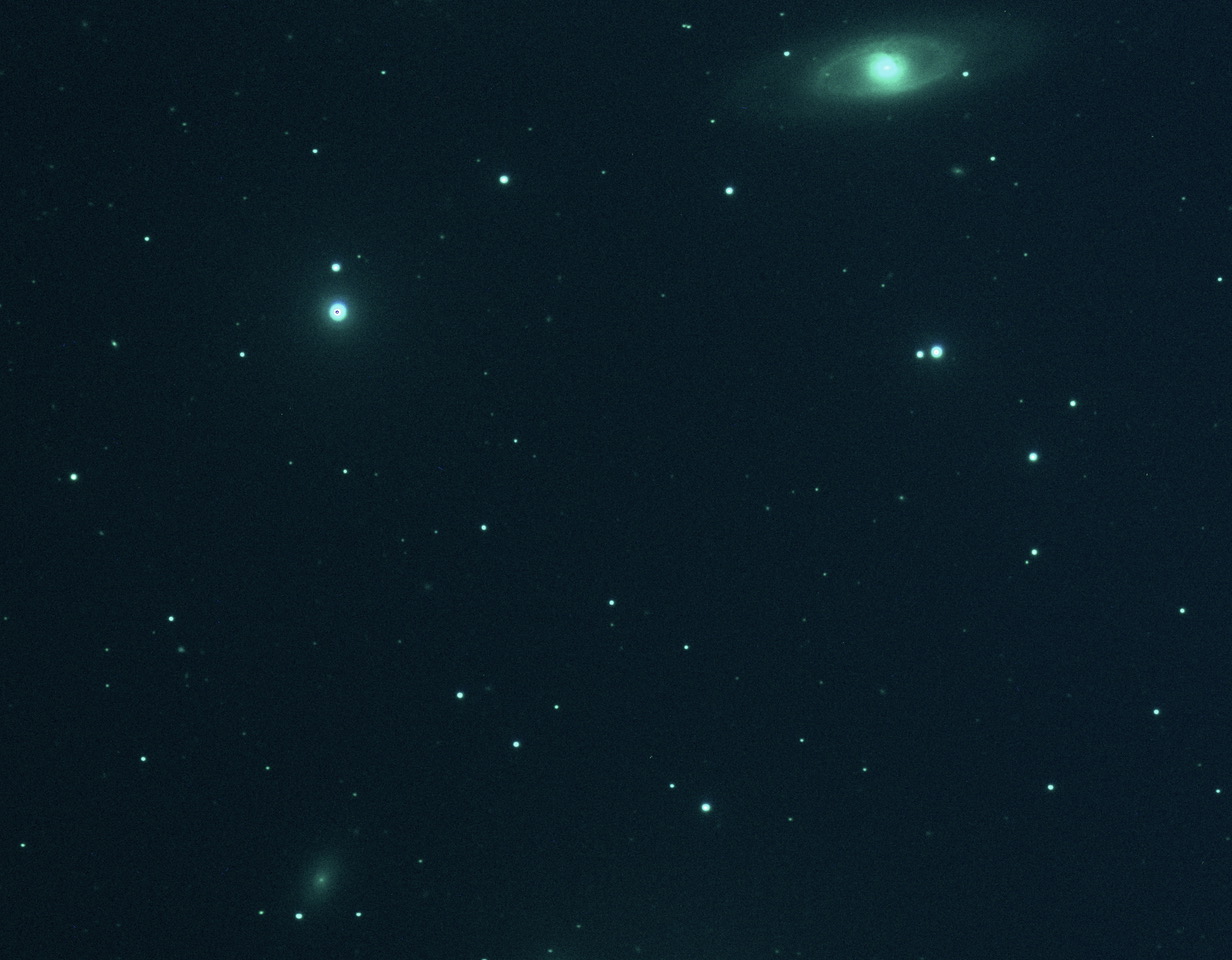}
\bigskip
\caption{ The NGC\,4274 galaxy can be seen in the top right corner of this colour composite image obtained from first light observations of the ILMT.}
\label{fig:Firstlight}
\end{figure}

The University of \LiU\ initiated the 4-m ILMT project and AMOS manufactured its bowl and mechanical structure. The 4-m ILMT was installed at Devasthal site in the year 2018 with the infrastructural support provided by ARIES. Fig.\,\ref{fig:ILMT} shows a top view of the mercury mirror where the mylar cover, preventing the formation of wavelets on the mercury surface, is clearly seen. The telescope received its first light on April 29, 2022. During the first commissioning phase of the telescope in April-May, 2022, photometric observations in $g, r$ and $i$ Sloan filters were acquired. A colour composite image of a small portion of the sky is shown in Fig.\,\ref{fig:Firstlight}. In this image, the NGC\,4274 galaxy can be seen in the top right corner. See \cite{surdej2022} and \cite{ 2022JAI....1140003K} for further details. Based on the first-light ILMT observations, 14 posters were presented during the \third\ BINA workshop. The preliminary results clearly demonstrate the significance of this low-cost telescope for a wide variety of celestial objects including transients and space debris. The 4-m ILMT in co-ordination with the 3.6-m DOT will strengthen the {\bf Indo-Belgian} co-operation significantly. The efforts of Prof. Jean Surdej from the University of \LiU\ , Belgium toward the success of this project are highly valued.
 
\begin{figure}[h!]
\centering
\includegraphics[width=12cm]{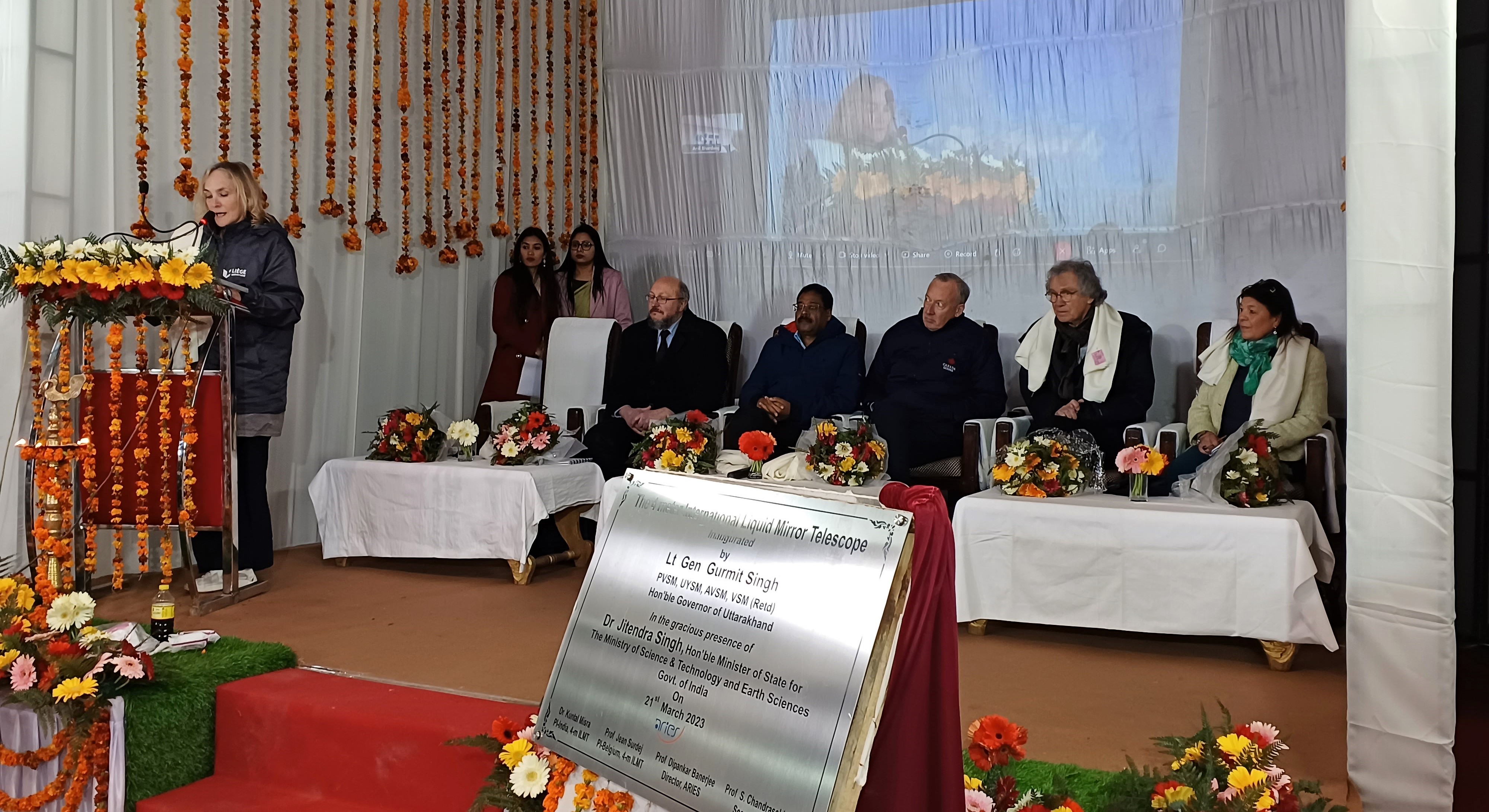}
\caption{A photo of the ILMT inauguration on 21 March 2023 at Devasthal is shown. }
\label{fig:ILMT_Ina}
\end{figure}

 On March 21, 2023, the ILMT was inaugurated jointly online by Lt. Gen. (Retd.) Gurmit Singh (honourable Governor of Uttarakhand) and  Dr. Jitendra Singh (honourable Minister of State for Science \& Technology and Earth Sciences, Govt. of India) in adverse weather conditions. To commemorate this momentous occasion, several dignitaries both from India and Belgium were present either online or in-person at the Devasthal Observatory. Dr. S. Chandrasekhar (secretary of DST, Govt. of India) and Prof. A. S. Kiran Kumar (former secretary of the Department of Space, Govt. of India) participated online in the function. The event was graced by the physical presence of His Excellency Mr. Didier Vanderhasselt (ambassador of Belgium in India), Prof. Anne-Sophie Nyssen (rector of the University of \LiU\,, Belgium), Dr. Ronald van der Linden (director of the ROB, Belgium), Dr. S.K. Varshney (head of the international division, DST, Govt. of India), Prof. Jean Surdej (project investigator of the ILMT from University of \LiU\,, Belgium), Mrs. Brigitte Decadt (BELSPO, Govt. of Belgium), and Prof. Dipankar Banerjee (director of ARIES, Nainital). Fig.\,\ref{fig:ILMT_Ina} shows a photo of the ILMT inauguration function. Congratulatory messages by Shri Pushkar Singh Dhami (honorable chief minister of Uttarakhand), Shri Ajay Bhatt (minister of state, Govt. of India; member of parliament, Nainital), and Prof. P. C. Agrawal (chairman of the governing council, ARIES) were portrayed during the inaugural event. The dignitaries addressed this watershed event and emphasized the astrophysical significance of the 4-m ILMT. The global astronomical community, thus, witnessed this historical event. On this occasion, a public talk titled {\it The 4-m International Liquid Mirror Telescope: A Short History} was delivered by Prof. Jean Surdej. The ILMT inaugural function ended with a welcome reception hosted by ARIES at the Country Inn Nature Resort, Bhimtal. 
 
\section{Accomplishments of BINA}
\label{accomp}
To achieve the objectives of BINA, three workshops were conducted along with a number of academic exchange work visits. They are briefly discussed in the next sub-sections. The scientific outputs of BINA are summarized in the remaining part of this section. All these indicate that the network has been successful but needs further strengthening by both countries. 

\subsection{ Workshops, work visits and training of next generation under BINA}

Three BINA workshops have been organised so far. ARIES hosted the $1^{\rm st}$ BINA Workshop on {\it Instrumentation and Science with the 3.6-m DOT and 4.0-m ILMT telescopes} at Nainital during 15-18 November 2016 \citep{2018BSRSL..87....1D}. Over 100 academicians from 8 countries (India, Belgium, Russia, Japan, China, South Africa, Thailand and Taiwan) participated in the workshop. The ROB organised the $2^{\rm nd}$ BINA Workshop in Brussels during 9-12 October, 2018. The theme of the workshop was {\it BINA as an Expanding International Collaboration} \citep{2019BSRSL..88....1D}. A total of 65 scientists from 7 countries (India, Belgium, France, Germany, Russia, Spain and Thailand) participated. The ROB also organized a one-day workshop on 8 October, 2018 during the work visits of three Indian scientists while an online ILMT workshop was organised by ARIES from June 29 to July 1, 2020 during the pandemic. The \third\ BINA workshop was hosted by ARIES from March 22 to 24, 2023 at the Bhimtal campus of the Graphic Era Hill University. Further details of this workshop are provided by \citet{Joshi2023} while its summary is given by \citet{semenko2023}. Peer-reviewed manuscripts of all three BINA workshops are published online in the Bulletin de la Soci\'{e}t\'{e} Royale des Sciences de \LiU.

Apart from above mentioned three joint international workshops, twelve work visits by researchers from both sides were conducted. Mutual visits were beneficial to strengthen the scientific activities and train the next generation of young PhD students. As a part of the human resources programme, about a dozen PhD students were/are jointly supervised by Indian and Belgian scientists. Two PhD students, namely Brajesh Kumar from ARIES, Nainital and Bikram Pradhan from Indian Institute of Space science and Technology (IIST), Thiruvananthapuram were trained for the ILMT project. The University of \LiU\,, Belgium awarded PhD degrees to them in November 2014 and January 2020, respectively, under the joint supervision led by Prof. Jean Surdej. About a dozen of Master and PhD students from both sides were trained towards the analysis of observational data and interpretation of the results. To date, a total of 8 PhD students from various Indian institutes received 3-month Erasmus+ mobility grants for working at the University of \LiU, under a collaborative program coordinated by Micha\"el De Becker. One of them, Bharti Arora, is now working as a post-doctoral researcher funded by Wallonia-Brussels International at the University of \LiU\,, Belgium. 

 Supervision of PhD work of African students is one of the mandates of the DST International co-operation division. Under this program, Dr. Otto Trust from Mbarara University of Science \& Technology (MUST), Uganda, completed his PhD under the joint supervision of the Indian and Belgian PI of the BINA project while Ms. Dorothy Museo Mwanzia from University of Nairobi and Mr. Kampindi Felix from MUST, Uganda are registered for their PhD work under Indian and Belgian researchers.

\subsection{Key science results obtained under BINA projects}

The 3.6-m DOT is the first major project which benefited from the Indo-Belgian collaboration. Both optical and near infra-red (NIR) observations taken with this telescope reveal that images of sub-arc-sec resolution can be obtained for a significant fraction of photometric nights. They also indicate that the performance of the telescope is comparable with its peers located elsewhere in the world \citep{2017CSci..113..682O, 2018BSRSL..87...29K, 2019CSci..117..365S, 2020JApA...41...33S, 2022JApA...43...31S}. This observing facility has led to a good number of national and international collaborations including BINA \citep{2019BSRSL..88...19J}. In one such collaboration, \cite{2021AJ....162..257A} combined 3.6-m DOT NIR data and $X-$ray observations taken with the {\it AstroSat} Soft $X-$ray telescope to investigate formation mechanisms of the colliding-wind binary WR125 (WC$7+O9$III). Analyses of these new measurements yield a period of long-term light variations of 28--29 years and support the recurrence time of episodic dust production. So far, the 3.6-m DOT has contributed to over 100 publications and 6 PhD theses in a number of front-line galactic and extra-galactic astrophysical research areas, including optical follow-up of Giant Meter Radio Telescope (GMRT) and {\it AstroSat} sources \citep{2022InJHS...57..227S}. 

The second telescope built under this collaboration is the 4-m ILMT. It was successfully installed in 2018 \citep{surdej2022, 2022JAI....1140003K} and has started routine observations after its inauguration on March 21, 2023. It is a zenith-pointing optical observing facility at Devasthal Observatory. The ILMT performs multi-band optical imaging of a narrow strip ($\sim$22 arcmin) of the sky. Software pipelines for the astrometric calibration, using the astrometric catalogues published in \cite{2022JAI....1140001D} and \cite{2020JApA...41...22M}, and the photometric calibration are in the development and testing phase \citep{2022JApA...43...10K}. Other than transient astronomy, the ILMT survey will detect and characterize the space debris which is present at altitudes ranging from low Earth orbits (LEO - an altitude of 2000 km or less) to geosynchronous orbits. It consists primarily of expired spacecraft, rocket stages, separation devices, and products of the collision or breakup of satellites. The study of their size distribution is an important input to risk analysis for current and future space missions. \cite{pradhan2019} investigated these aspects with the 1.3-m DFOT telescope and found that the 4-m ILMT photometric survey may provide detections of objects having diameters as small as 3 cm in a LEO. The ILMT team has published over a dozen papers and contributed to two PhD theses, so far.

Optical characterization and Radial velocity monitoring with Belgian and Indian Telescopes (ORBIT) is an ongoing project under BINA to study the exoplanets and low-mass eclipsing binary stars. For this, the photometric observations are taken from ARIES (Nainital; India) \citep{2022JAI....1140004J} while the spectroscopic observations are taken with the Hanle Echelle Spectrograph at the 2-m Himalayan Chandra Telescope (Hanle; India) \citep{2018SPIE10702E..6KS} and the High Efficiency and high-Resolution Mercator Echelle Spectrograph (HERMES) mounted at the 1.2-m Mercator Telescope, La Palma; Canary Islands in Spain \citep{2011A&A...526A..69R}. The combined observations are used to estimate basic parameters of the binary systems (e.g. orbital periods, masses of the components, radii) \citep{2023MNRAS.521..677P}. 

At the beginning of the year 2000, a bilateral research project entitled ``The Nainital-Cape survey'' was jointly initiated by the scientists from the ARIES in India and from the South African Astronomical Observatory in South Africa. It was supported by DST (Govt. of India) and the National Research Foundation (Govt. of South Africa). This was one of the largest ground-based photometric surveys ever taken with the aim to search and study the photometric variability in a large sample of chemically peculiar (CP) stars (\citealt{2000ashoka}, \citealt{2001martinez}; \citealt{joshi2003, 2006joshi, joshi2009, joshi2010, joshi2012,joshi2016,joshi2017}). By combining time-series of ground- and space-based photometry with high-resolution spectroscopy of a set of samples observed using multiple telescopes around the globe, \cite{joshi2022} discovered a new heartbeat system (HD\, 73619), for which no pulsation signatures are seen. Such works not only aim to study the stellar structure and atmospheres of CP stars in the presence of magnetic fields, inhomogeneities (such as spots), and tidal interactions but also demonstrate the importance of international (13 countries) co-operation led by Indians and Belgians.

Another active area of the collaboration has been the photometric and spectroscopic study of eclipsing binaries with the
aim to understand the formation mechanism of these stars at different stages of their evolution. For example, 56 UMa is a wide binary system containing a CP red giant with a faint companion. To unravel the nature of its faint companion, \cite{2023A&A...670L..14E} revisited the orbital parameters of the system and carried out a detailed spectral analysis including high-resolution HERMES spectra. This study indicates that unseen component in 56 UMa has a mass of 1.31$\pm 0.12 
 M_\odot$, which is compatible with the mass of both a white dwarf (WD) and a neutron star. However, some observations are in favor of the latter interpretation.

Magnetic cataclysmic variables (MCVs) are interacting semi-detached binaries consisting of a magnetic WD as the primary and a Roche lobe filling star as the secondary. The magnetic field strength of the WDs divides the MCVs in two sub-classes namely intermediate polars and polars. Based on detailed optical and $X-$ray timings and a spectral study of two candidate MCVs, namely 1RXS J174320.1-042953 and YY Sex, \cite{2023MNRAS.521.2729R} conclude that these MCV candidates belong to the polar sub-class of MCVs.

Gravitational microlensing by compact objects in lensing galaxies is a known tool for probing the structure of distant quasars on sub-parsec scales. For this purpose, \cite{2020A&A...633A.101H} used the Very Large Telescope of the European Southern Observatory to obtain spectropolarimetric observations of the two images of the broad absorption line (BAL) quasar SDSS $J081830.46+060138.0$ at redshift $z=$2.35. These observations indicate that the underlying source is actually gravitationally lensed and not a binary quasar. Detection of an absorption system at z=1.0065$\pm$0.0002 might reveal the lensing galaxy. The authors also found that this is the second BAL quasar in which an extended source of the rest-frame ultraviolet continuum is found. This study provides constraints on the BAL flow and finds that the outflow is seen with a nonzero onset velocity stratified according to ionization.

\cite{2020A&A...642A.136B} used the GMRT observations at frequencies 325 and 610 MHz, for a detailed study of 11 early-type stars including both WR and O-type systems, located in $\sim$ 15 degree$^2$ area of the sky centred on the Cygnus region and identified two additional particle-accelerating colliding-wind binaries, namely Cyg OB2 12 and ALS 15108 AB. This project was part of a collaborative network led by Mich\"el De Becker from Belgium (PANTERA-Stars) with Indian scientists from National Center of Radio Astronomy, Pune and IIST, Thiruvananthapuram who are making use of the GMRT observing facility for the multi-wavelength studies of systems consisting of massive stars active at accelerating particles up to relativistic velocities.

\section{Conclusion and summary}
Especially during the last decade, the Indo-Belgian co-operation and BINA activities have been very productive and scientifically rewarding. This fruitful co-operation helped and lead to the installation of two 4-m class telescopes at Devasthal Observatory, being the largest optical ones in India. Not only a good number of next-generation young minds have been trained but also over 50 joint papers have been published  in peer-reviewed high-impact international journals (including 25 in main journals and 26 in the proceedings of the $1^{\rm st}$ and $2^{\rm nd}$ BINA workshops). As a result, BINA has grown into a household name in the astronomical community within Belgium and India that aims to stimulate all forms of Indo-Belgian research in astronomy and space sciences. All the network activities, supported by the funding agencies DST (at the Indian side) and BELSPO (at the Belgian side), have created a strong foundation for its future. Latest addition to these activities is the creation of the network Belgo-Indian Projects on precision Astronomical spectroscopy for Stellar and Solar system bodies (BIPASS) in 2022. It is led by Prof. Shashikiran Ganesh from Physical Research Laboratory, Ahmadabad (India) and Dr. Laurent Mahy from ROB, Brussels (Belgium). There are 5 Indian and 4 Belgian partners in the BIPASS project. Research in atmospheric sciences is one of the areas of common interest between Indian and Belgian scientists that has not been explored yet, but BINA intends to include it in their domain of activities too. All these portray a bright future for BINA co-operation. 

\begin{acknowledgments}

Constructive suggestions made by Prof. Edward Guinan, reviewer of the manuscript, are thankfully acknowledged. This work is supported by the Belgo-Indian Network for Astronomy and astrophysics (BINA), approved by the International Division, Department of Science and Technology (DST, Govt. of India; DST/INT/BELG/P-09/2017) and the Belgian Federal Science Policy Office (BELSPO, Govt. of Belgium; BL/33/IN12). The author acknowledges financial support provided by the organizing committee of the \third\ BINA workshop. He is also grateful to the National Academy of Sciences, Prayagraj, India for the award of Honorary Scientist position, the Alexander von Humboldt Foundation (Germany) for the award of long-term group research linkage program, and director of the Indian Institute of Astrophysics; Bangalore, India for providing infrastructural support during this work. Valuable inputs provided by Drs. Peter De Cat, Mich\"el De Becker, Santosh Joshi and Kuntal Misra are thankfully acknowledged.   

\end{acknowledgments}

\begin{furtherinformation}

\begin{orcids}
\orcid{0000-0003-4973-4745}{Ram} {Sagar}

\end{orcids}

\begin{conflictsofinterest}
The authors declare no conflict of interest.
\end{conflictsofinterest}

\end{furtherinformation}

\bibliographystyle{bullsrsl-en}

\bibliography{extra}

\end{document}